\documentclass[prd,twocolumn,amsmath,amssymb,floatfix,superscriptaddress]{revtex4-1}
\usepackage{bm}
\usepackage{amsmath}
\usepackage{epsfig}
\usepackage{color}
\usepackage{natbib}
\usepackage{graphicx}
\usepackage{hyperref}
\usepackage{ifthen}
\usepackage{xstring}
\usepackage{graphicx}

\newcommand{\curv}{{\cal R}}

\definecolor{darkgreen}{cmyk}{0.85,0.2,1.00,0.2} 
\definecolor{purple}{cmyk}{0.5,1.0,0,0}

\begin{document}
\title{Generalized Slow Roll  for Tensors}

\author{Wayne Hu}
\affiliation{Kavli Institute for Cosmological Physics, Department of Astronomy \& Astrophysics, Enrico Fermi Institute, University of Chicago, Chicago IL 60637}

\date{\today}

\begin{abstract}
\baselineskip 11pt
The recent BICEP2 detection of  degree scale CMB B-mode polarization, coupled with a deficit of observed power in large angle temperature anisotropy, suggest that the
slow-roll parameter $\epsilon_H$, the fractional variation in the Hubble rate per efold, is
both relatively large and may evolve from an even larger value on scales greater than the horizon at recombination.  The relatively large tensor contribution implied also requires
finite matching features in the tensor power spectrum for any scalar power
spectrum feature proposed to explain anomalies in the temperature data.  
We extend the generalized slow-roll approach  for computing power spectra, 
appropriate for such models where 
the slow-roll parameters vary, to tensor features  where scalar features are large.  This approach 
also generalizes the tensor-scalar consistency relation to be between the ratio of tensor and scalar sources and features in the two power 
spectra.   Features in the tensor spectrum are generically suppressed by $\epsilon_H$ relative those in the scalar spectrum and by the smoothness of the Hubble rate, which must obey covariant conservation of energy, versus its derivatives.   Their detection in near future CMB data would
 indicate a fast roll period of inflation where $\epsilon_H$ approaches order unity, allowed but not required by inflationary explanations of temperature anomalies.
\end{abstract}

\maketitle

\section{Introduction} \label{sec:intro}

The recent detection of inflationary tensor modes from $B$-mode polarization of the cosmic microwave background (CMB)
by the BICEP2 experiment imply a large scalar to tensor ratio $r=0.2^{+0.07}_{-0.05}$ \cite{Ade:2014xna} while also hinting at
a violation of ordinary slow-roll prediction of a nearly scale-free curvature power spectrum.  
The latter is associated with the deficit rather than
increment in the large angle Planck temperature power spectrum (e.g.\ \cite{Contaldi:2014zua,Miranda:2014wga,Abazajian:2014tqa,Hazra:2014jka,Bousso:2014jca}) which would imply
$r< 0.11$ (95\% CL) in the standard cosmological constant, cold dark matter $\Lambda$CDM model
\cite{Ade:2013uln}.  Other hints of transient violations include glitches in the low multipole temperature spectrum \cite{Peiris:2003ff}
and high frequency oscillations in the high multipole temperature spectrum \cite{Flauger:2009ab,Adshead:2011jq,Ade:2013uln}.

Such violations also leave imprints on the tensor power spectrum.   Generically tensor features require changes in the Hubble
rate $H$ during inflation while scalar features only require changes in its derivatives or the inflaton sound speed.  Order unity features 
would require order unity changes in $H$ and hence typically an interruption of inflation.   Just like the tensor-scalar ratio itself, tensor features
are suppressed relative to scalar features by a factor of the slow-roll parameter $\epsilon_H$,
the fractional evolution of the Hubble parameter per efold
 (e.g.~\cite{Hamann:2007pa,Hazra:2010ve}).   On the other hand the relatively large
$r$ implied by BICEP2 now requires a finite rather than infinitesimal $\epsilon_H$.   It is therefore timely
to develop general tools for the prediction of tensor features and study their consistency relation to scalar features.

 The generalized slow-roll (GSR) approach was introduced in Ref.~\cite{Stewart:2001cd,Choe2004,Gong:2004kd} to compute  power spectra for
 models where the ordinary slow-roll parameters vary strongly with time but the background remains close to de Sitter.   
 It was subsequently extended for order unity scalar power spectrum features  \cite{Dvorkin2010}, for general
 single-field inflation \cite{Hu2011}  and for the curvature bispectrum  \cite{Adshead:2011bw,Adshead:2013zfa}.
 
 Here we further develop these techniques for  tensor power spectrum features 
  and explore
 their relation to scalar power spectrum features in the limit that the latter are large.   
 In \S \ref{sec:GSR}, we extract the
 common mathematical structure of the GSR approach that is applicable to both
 scalars and tensors.   In \S \ref{sec:TS}, we highlight the consistency relation,
 similarities and differences between scalar and tensor features.   We provide
 examples in \S \ref{sec:examples} motivated by anomalies in the temperature power spectrum and discuss these results in \S \ref{sec:discussion}.

\section{Generalized Slow Roll}
\label{sec:GSR}

We begin by extracting the mathematical content of the GSR technique
which is common to both scalar and tensor modes.  As we shall see in \S \ref{sec:TS}, scalar curvature
fluctuations and tensor gravitational waves satisfy a common evolution equation 
for their modefunctions $y$
\begin{equation}
{d^2 y \over dx^2} + \left( 1 - {2 \over x^2} \right) y = \left( f'' - 3  \frac{f'}{f}\right) \frac{y}{x^2} ,
\label{eqn:yeqn}
\end{equation}
from Bunch-Davies initial conditions 
\begin{equation}
\lim_{x\rightarrow \infty} y = e^{i x}.
\end{equation}
Here $'=d/d\ln x$ where $x$ is the time variable which for scalars will be associated with the sound horizon
and for tensors the horizon; for both $x$ runs from infinity to zero as inflation progresses. 
Likewise scalars and tensors will have a different
 source of excitations from pure de Sitter conditions characterized by the time evolution of the function $f$. 
We shall see that $f$ in each case is related to the slow-roll parameters.

The GSR technique assumes that the excitations in the modefunctions
are small rather the derivatives of $f$ itself, thus allowing the latter to evolve strongly due to inflationary features.
Under this assumption, the modefunction equation can be solved iteratively by first
setting the rhs of Eq.~(\ref{eqn:yeqn}) to zero to obtain the de Sitter mode functions
\begin{equation}
y_0 = \left( 1 + {i \over x} \right) e^{i x} ,
\end{equation}
then replacing $y \rightarrow y_0$ in the rhs  to solve for
the first order correction $y_1$.    This process may be repeated to arbitrary order.   
Iteration to second order yields for the superhorizon power
\begin{align}
\Delta^2 &\equiv
\lim_{x \rightarrow 0} \left| { \frac{x y}{f}  }\right|^2 \nonumber\\
& \approx e^{I_0}\left[ \left(
1 + \frac{1}{4}I_1^2 + \frac{1}{2} I_2 \right)^2 + \frac{1}{2} I_1^2 \right].
\label{eqn:GSRpower}
\end{align}
Here the leading order term is
\begin{equation}
I_0 = G(\ln x_{\rm min}) + \int_{x_{\rm min}}^\infty {d x\over x} W(x) G'(\ln x) ,
\label{eqn:I0}
\end{equation}
where $x_{\rm min}\ll 1$ and
\begin{equation}
G = - 2 \ln f    + {2 \over 3} (\ln f )'  
\label{eqn:G}
\end{equation}
with $' = d/d\ln x$.   
The window function 
\begin{equation}
W(x) = {3 \sin(2 x) \over 2 x^3} - {3 \cos (2 x) \over x^2} - {3 \sin(2 x)\over 2 x} 
\end{equation}
determines how the deviations from de Sitter freeze out.
Note that this construction introduces $G'$ in place of $f'' - 3f'/f$ for these deviations so as
to preserve the
constancy of $ | xy /f|$ above the horizon order by order, regardless of the
size of $G'$.    Furthermore, in the ordinary slow-roll approximation, where $G$ is
taken to be nearly constant $| x y / f|^2 \rightarrow e^{G} \approx 1/f^2$, consistent
with de Sitter modefunctions where $y_0 \rightarrow i/x$.   Thus we shall see that
$1/f^2$ is associated with the scalar and tensor power spectra $\Delta^2$ in the ordinary slow-roll
approximation.

The second order corrections are
\begin{eqnarray}
I_1 &=& { 1\over \sqrt{2} } \int_0^\infty {d x \over x} G'(\ln x) X(x) , \nonumber\\
I_2 &=& -4 \int_0^\infty { d x \over x } [ X + {1\over 3} X' ] {f' \over f} F_2(x) ,
\end{eqnarray}
with 
\begin{equation}
X(x) = {3 \over x^3} (\sin x- x \cos x)^2 ,
\end{equation}
and
\begin{equation}
F_2(x) = \int_x^\infty {d  u \over  u^2} {f' \over f}.
\end{equation}
The validity of the GSR expansion can be checked by calculating these corrections 
and ensuring that they are small compared with the leading order term.

\section{Tensors vs.\ Scalars}
\label{sec:TS}

Now let us review the application of the GSR technique to scalar or comoving curvature fluctuations $\curv$ and then apply it to tensors.
The curvature modefunctions obey the evolution equation
\begin{equation}
\frac{d}{d\eta} \left( \frac{a^2 \epsilon_H}{c_s^2} \frac{ d \curv}{d\eta} \right) + a^2 \epsilon_H 
 k^2 \curv =0 ,
 \label{eqn:Reqn}
\end{equation}
where $c_s$ is the propagation or sound speed of the scalar  fluctuations
and can differ from unity if there are non-canonical kinetic terms in the inflaton Lagrangian.
Deviations from a pure de Sitter expansion $H=$ const. are characterized by the
slow-roll parameter
\begin{equation}
\epsilon_H  = - \frac{d \ln H}{d  N}.
\end{equation}

The curvature modefunction equation (\ref{eqn:Reqn}) can be mapped onto the GSR modefunction equation (\ref{eqn:yeqn}) with the association \cite{Hu2011}
\begin{equation}
\curv \equiv \sqrt{ \frac{2 \pi^2}{k^3} } \frac{ x_\curv y_\curv}{f_\curv}.
\label{eqn:Rk}
\end{equation}
To distinguish scalar and tensor quantities, we  append a subscript $\curv$ to GSR variables
involving scalars.  Here
$x_\curv= k s$ where 
\begin{equation} 
s(N) = \int_N^0 d\tilde N \frac{c_s}{a H}
\end{equation}
is the sound horizon at an efold $N=\ln(a/a_{\rm end})$ from the end of inflation and
\begin{align}\label{eqn:fdef}
	f_\curv^2 & =  \frac{8 \pi^2 \epsilon_H c_s}{H^2} \left( \frac{a H s}{c_s} \right)^2.
\end{align}

The GSR source for 
Eq.~(\ref{eqn:I0}), $G_\curv'$, is a function of $\ln s$ independent of the wavenumber
and $' = d/d\ln x_\curv = d/d\ln s$.   
The curvature power spectrum is defined as
\begin{equation}
\Delta_\curv^2 = \lim_{ks \rightarrow 0}\frac{k^3}{2\pi^2} |\curv|^2 = \lim_{x_\curv \rightarrow 0}
\left|\frac{ x_\curv y_\curv}{f_\curv} \right|^2, 
\end{equation}
and so Eq.~(\ref{eqn:GSRpower}) says that to leading order
\begin{equation}
\ln \Delta_\curv^2(k)
 \approx G_\curv(\ln s_{\rm min}) + \int_{s_{\rm min}}^\infty {d s\over s} W(k s) G_\curv'(\ln s).
\end{equation}
Note that in the ordinary slow-roll approximation 
\begin{equation}
G'_\curv \approx {\rm const.} \equiv  1- n_S \qquad {\rm (SR)},
\end{equation}
and
\begin{equation}
\ln \Delta_\curv^2(k)  \approx G_\curv(\ln s_{\rm min})   +(n_S-1) [\ln (ks_{\rm min})-C]
\label{eqn:ns}
\end{equation}
consistent with a tilt of $n_S-1 = d\ln \Delta_\curv^2/d\ln k$ where $C= 7/3-\gamma_E-\ln 2$
with $\gamma_E$ as the Euler-Mascheroni constant.

Next consider the tensor fluctuations which represent gravitational wave amplitudes $h_{+,\times}$.   Their modefunction equation, for either polarization and wavenumbers much smaller than the curvature scale, is 
\begin{equation}
\frac{d^2 h}{d \eta^2} + \frac{2}{a} \frac{d  a}{d\eta}\frac{d h}{d\eta} + k^2 h =0,
\label{eqn:h}
\end{equation}
where 
\begin{equation}
\eta(N) = \int_N^0 d\tilde N \frac{1}{a H}
\end{equation}
is the horizon measured from the end of inflation.   The  gravitational wave modefunction is
related to $y_h$ through canonical normalization
\begin{equation}
h \equiv \sqrt{ \frac{2 \pi^2}{k^3} } \frac{ x_h y_h}{f_h},
\label{eqn:hk}
\end{equation}
with \cite{Gong:2004kd}
\begin{align}
	f_h^2 & =  \frac{2 \pi^2}{H^2} \left( {a H \eta} \right)^2 \propto (a\eta)^2,
	\label{eqn:fh}
\end{align}
 which transforms Eq.~(\ref{eqn:h}) into the modefunction equation Eq.~(\ref{eqn:yeqn})
with the association
\begin{equation}
 f_h'' - 3  \frac{f_h'}{f_h} = \frac{\eta^2}{a} \frac{d^2 a}{d\eta^2} -2 .
\end{equation}

Thus the tensor power spectrum in each polarization state is
\begin{equation}
\Delta_{+,\times}^2 = \lim_{k\eta \rightarrow 0}\frac{k^3}{2\pi^2} |h|^2 = \lim_{x\rightarrow 0}
\left|\frac{ x_h y_h}{f_h} \right|^2
\end{equation}
and Eq.~(\ref{eqn:GSRpower}) determines how deviations in the source function $G'_h$ freeze out.  Explicitly, to leading order
\begin{equation}
\ln \Delta_{+,\times}^2(k)
 \approx G_h(\ln \eta_{\rm min}) + \int_{\eta_{\rm min}}^\infty {d \eta\over \eta} W(k \eta) G_h'(\ln \eta),
 \label{eqn:hpower0}
\end{equation}
where in distinction to scalars $'=d/d\ln x_h= d/d\ln\eta$ and
\begin{equation}
G'_h = -2 (1- a H\eta) - \frac{2}{3} aH\eta \left( 1- aH\eta + a H\eta \,\epsilon_H\right).
\end{equation}
In the ordinary slow-roll approximation $a H \eta \approx 1+ \epsilon_H$ and hence
\begin{equation}
G'_h \approx 2 \epsilon_H = -n_T \qquad {\rm (SR)},
\label{eqn:Gptsr}
\end{equation}
which
yields
\begin{equation}
\ln \Delta_{+,\times}^2(k)  \approx G_h(\ln \eta_{\rm min})   +n_T [\ln (k\eta_{\rm min})-C]
\label{eqn:nt}
\end{equation}
 consistent with  $n_T= d\ln \Delta^2_{+,\times}/d\ln k$ as the tensor tilt.  Note that even beyond the constant $\epsilon_H$ approximation
\begin{equation}
(1-a H\eta)={\cal O}(\epsilon_H),
\end{equation}
and hence $G'_h = {\cal O}(\epsilon_H)$.

There are several similarities and difference between scalars and tensors in the context
of inflationary features that are worth highlighting.  In place of the direct consistency relation
between the tensor-scalar ratio and the tilt of the tensor power spectrum is a 
relationship between the GSR source functions
\begin{equation}
r_f \equiv 4\frac{f_{\curv}^2}{f_h^2} =  16 {\epsilon_H c_s} \left(\frac{s}{c_s\eta} \right)^2.
\label{eqn:rf}
\end{equation}
Recall that in the slow-roll approximation $1/f^2$ is directly related to the corresponding
power spectrum and $s \approx c_s\eta$ so that
\begin{equation}
r_f \approx \frac{4 \Delta^2_{+,\times}}{\Delta^2_\curv} \equiv r \approx 16\epsilon_H c_s\qquad {\rm (SR)},
\end{equation}
which in turn is related to the tilt of the tensor spectrum in slow roll 
through Eq.~(\ref{eqn:Gptsr}).
Beyond the ordinary slow-roll limit of constant $G'$, the consistency relation (\ref{eqn:rf}) still implies that the 
GSR $f$ sources, and hence power spectrum features, are related.  However their appearance in the respective power spectra
goes through the freezeout integrals over $G'$ rather than simply the tensor tilt. 

Freezeout occurs at different epochs for tensors than scalars.   For example
a feature at some common efold $N_s$ induces changes to the $\curv$ and $h$ modefunctions if
$k \gtrsim 1/\eta(N_s)$ for tensors but only if $k \gtrsim 1/s(N_s)$ for scalars.
Hence for $c_s\ll 1$, there is a range of
 wavenumbers where a feature can impact tensors while not affecting scalars which
 have already passed through the sound horizon.  
 
Next only features in the evolution of $H$ or equivalently $\epsilon_H$ affect tensors
whereas features in $c_s$ also affect scalars.   Moreover, features in $\epsilon_H$ 
impact scalars more than they do tensors as long as $\epsilon_H \ll 1$, the
requirement of a near de Sitter expansion which is at the heart of both the ordinary and
generalized slow-roll approximations.   Energy conservation guarantees that $H$ evolves
continuously and energy loss to the expansion can only occur on the efold time scale.

 The tensor GSR sources therefore  yield
 $O(\Delta \epsilon_H)$ fractional effects on top of the slow-roll power spectrum through 
 $G'_h$,
 whereas the scalar GSR sources produce $O(\Delta \epsilon_H/\epsilon_H)$ fractional
 effects.   Thus we generically expect tensor power spectrum features to be suppressed
 relative to scalar power spectrum features by at least a factor of $\epsilon_H$.

The BICEP2 result suggests that tensor to scalar ratio $r \approx 0.2$ and so $\epsilon_H
\sim 0.2/16 c_s$ during the slow-roll period.   For canonical sound speed models $c_s=1$ and we generically
expect features in the tensor sector to be percent level  for order unity scalar features.
For low sound speed models, the effect can be larger but only at the expense of
making $\epsilon_H$ larger which in turn limits the number of efolds that slow-roll
inflation can proceed.   Thus  tensor features that are comparable to scalar features typically  require a short duration fast roll period of inflation where $\epsilon_H$ itself reaches order unity.  While a truly fast roll period would also suppress scalar fluctuations on large 
scales and assist in reconciling the BICEP2 result with upper limits on $r$ from Planck,
we shall show in the next section it is not required to resolve this tension.

  \begin{figure*}[t]  
  \psfig{file=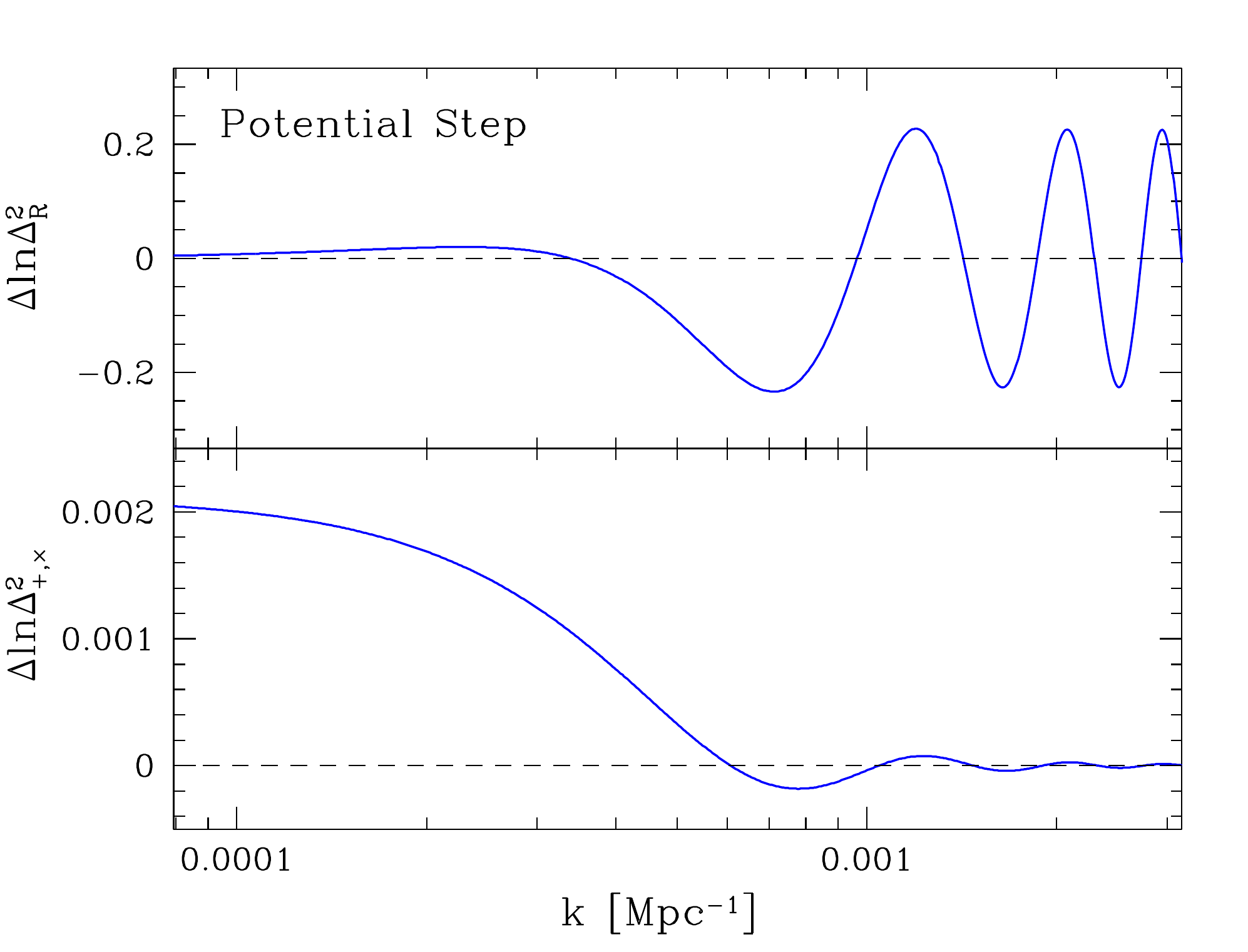, width=3.25in}
\psfig{file=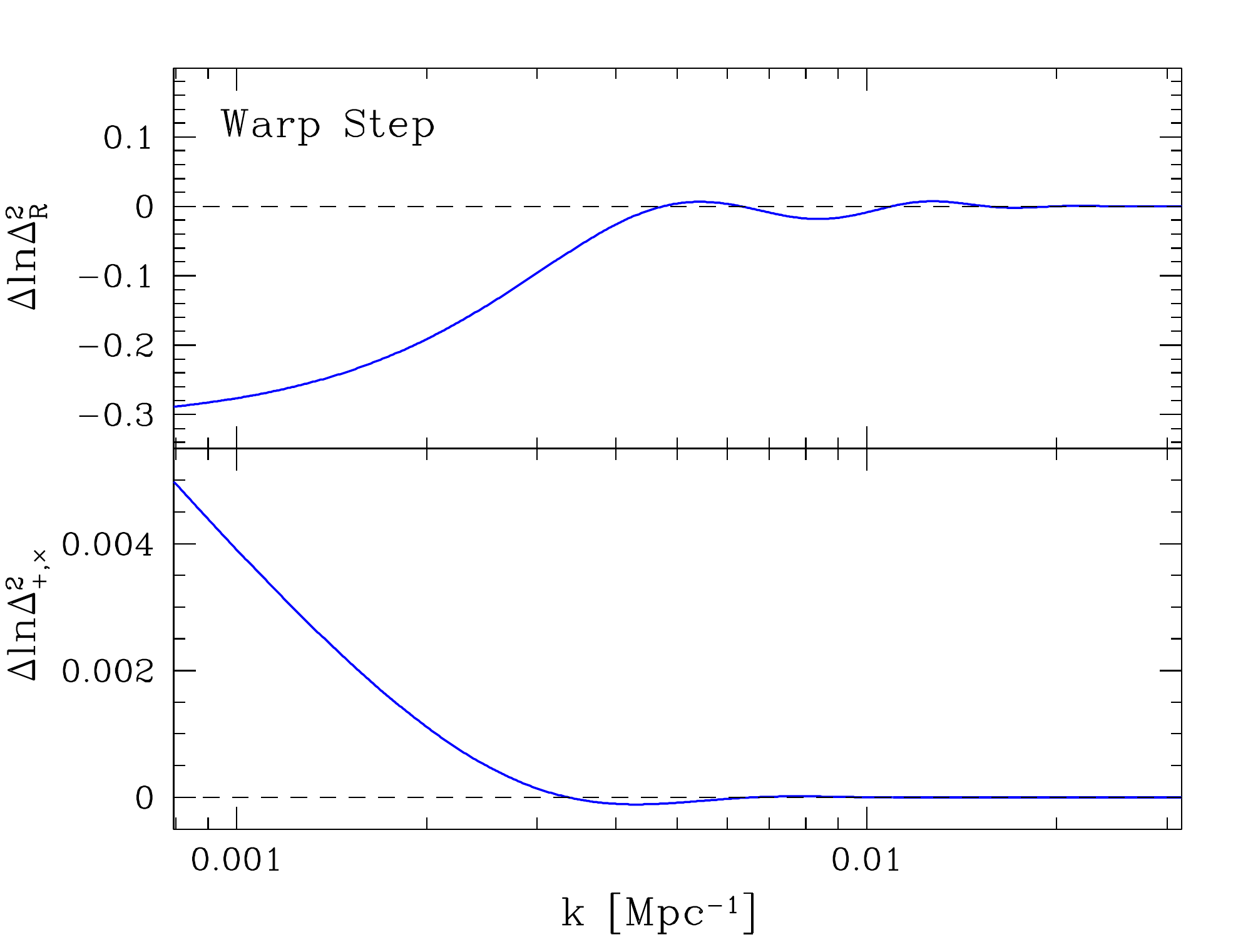, width=3.25in}
\caption{Examples of scalar and tensor features.   Left: extremely sharp potential step that
fits high frequency oscillations in the Planck temperature anisotropy data.   Right: warp step 
that fits the scalar suppression implied by the BICEP2 $B$-mode detection and the
Planck large angle temperature power spectrum.   In both cases, tensor features are
suppressed by a factor of $\epsilon_{H}\ll 1$ and oscillations are further suppressed by the
relative smoothness of the tensor source due to energy conservation. }
\label{fig:feature}	
\end{figure*}

\section{Examples}
\label{sec:examples}

Features in the scalar and tensor power spectra arise from the evolution of
$\epsilon_H$ and $c_s$ which excite the scalar and tensor modefunctions from their
de Sitter forms. 
We have seen in the previous section that tensor features are generically small so long as the background is nearly de-Sitter or $\epsilon_H \ll 1$, even if the evolution of $\epsilon_H$ is
strong enough to induce large scalar features $\Delta\epsilon_H/\epsilon_H = {\cal O}(1)$.   In this section we quantify this expectation
with concrete models that are motivated by features in the observed CMB temperature
power spectrum \cite{Miranda:2013wxa,Miranda:2014wga}.

These models involve sharp steps in for example the potential or warp factor of DBI
inflation, including its $c_s \rightarrow 1$ limit of a canonical scalar field inflaton.
Following 
\cite{Miranda:2012rm,Miranda:2013wxa},
the evolution of $\epsilon_H$ and $c_s$  is generically quantified by the change in their
values from before (``$b$") to immediately following the step (``$i$"), as determined by energy conservation, and then their decay back to their inflationary attractor values (``$a$") several
efolds after the step.   It is convenient to normalize the former quantities to the values
$\epsilon_{Ha}$ and $c_{sa}$ after the step leaving the free parameters
\begin{eqnarray}
e_b &=& \frac{\epsilon_{Hb}}{\epsilon_{Ha}}, \qquad
e_i = \frac{\epsilon_{Hi}}{\epsilon_{Ha}} ,
\nonumber\\
c_b &  =  &\frac{c_{sb}}{c_{sa}},  \qquad\ c_i  =  \frac{c_{si}}{c_{sa}} .
\label{eqn:csepsHmodel}
\end{eqnarray}
Let us represent the step itself with the function $F$, which takes on the value
$-2$ before the step $N<N_s$ and $0$ after the step $N>N_s$.   Then the 
evolution across the step becomes
\begin{align}
\label{eqn:parammodel}
 \frac{c_s}{c_{sa}}(N) =&1+ \frac{1-c_b}{2} F + \frac{c_i-1}{2} (F+2) e^{3(N_s-N)}, \\
\frac{\epsilon_{H}}{\epsilon_{Ha}}(N) =&1+ \frac{1-e_b}{2} F + \frac{e_i-1}{2} (F+2) e^{3(N_s-N)}.\nonumber
\end{align}
$G_{\curv}'$ and $G_{h}'$ can then be computed by taking derivatives and integrals
of these fundamental quantities.   By defining $H$ through the integral of $\epsilon_H$ we
guarantee energy conservation which is crucial in establishing its continuous evolution in the
presence of the step.

To leading order, the change in the scalar power spectrum due to the step is \cite{Miranda:2012rm}
\begin{align} \label{eqn:scalarstep}
	 \Delta \ln \Delta_{\mathcal{R}}^2 = &
	 C_{1} W(k s_s) + C_{2} W'(k s_s) 
 + C_{3} Y(k s_s),
	 \end{align}
where 
\begin{align}
	Y(x) &= \frac{6x\cos(2x) + (4x^2-3) \sin(2x)}{x^3},
\end{align}
and \cite{Miranda:2013wxa}
\begin{align} \label{eqn:CW}
    C_{1} &= -\ln c_b e_b , \nonumber\\
    C_{2} &= -\frac{2}{3} \frac{c_i-c_b}{c_i + c_b}  + \frac{2}{3} \frac{e_i-e_b}{e_i+e_b},\nonumber\\
    C_{3}& = 2\frac{(1-c_b) +(c_i-1)/4}{c_i+c_b},
\end{align}
and we have assumed $\epsilon_{Ha} \ll 1$.   The $W$ term represents a step in
the scalar power spectrum due to change in $\epsilon_H c_s$.   The $W'$ term
produces constant amplitude ringing in the power spectrum as the transfer of
an infinitely sharp step.   The $Y$ term alters the spectrum around $k s_s=1$
by enhancing the sharpness of the step.

For the tensor spectrum, a similar series of calculations yields
\begin{align}
	\Delta \ln \Delta_{+,\times}^2 = &
	 B_{1} W(k \eta_s)   + B_{2} V(k\eta_s) + B_{3} Y(k \eta_s),
 \label{eqn:tensorstep}
	 \end{align}
	 where 
\begin{align}
V(x) &= \int_x^\infty\frac{d \tilde x}{\tilde x} W(\tilde x) \nonumber\\
& = 	 \frac{ (1+ x^2) \sin (2 x)-2 x \cos (2 x) }{2 x^3}-{\rm Ci}(2 x) ,
\end{align} 
with Ci as the cosine integral and
\begin{align} \label{eqn:Bi}
    B_{1} &=\frac{8}{3} \left[(1-e_b)+ \frac{1}{4}(e_i-1)\right]\epsilon_{Ha}  , \nonumber\\
    B_{2} &= 2(e_b-1) \epsilon_{Ha}
 ,\nonumber\\
    B_{3} &= \left[  (1-e_b)+\frac{1}{4}(e_i-1) \right]\epsilon_{Ha}.
     \end{align}
Note that in comparison with the $C_i$ scalar amplitudes, the $B_i$ tensor amplitudes
are all suppressed by a factor of $\epsilon_{Ha}$ as discussed in the previous section.

The new $V$ term here has the limit
\begin{align}
\lim_{x\rightarrow 0} V(x) = C  -\ln x ,
\end{align}
where $C$ was defined in Eq.~(\ref{eqn:ns}) and induces an effect similar to a change in the tensor tilt since $d V/d\ln k = dV/d\ln x$.
Indeed  the difference between $\epsilon_{Hb}$ and $\epsilon_{Ha}$ represented by a finite $e_b-1$
 changes the tensor tilt according to Eq.~(\ref{eqn:Gptsr})
in the slow-roll attractors away from step.  
    While it induces a similar effect on the
scalar spectrum, the other $C_i$ effects are zeroth order effects in $\epsilon_{Ha}$ whereas
all tensor $B_i$ effects begin at first order.   

Finally for a finite width step, the oscillatory features in both spectra are damped at 
high wavenumbers due to the fact that the $W$ window function oscillates many times
while the step is being traversed.   In the sharp step limit where this duration is still
much less than an efold $x_d \approx 1/\delta N \gg 1$, the $\Delta\ln \Delta^2$ power
spectrum deviations of Eq.~(\ref{eqn:scalarstep}) and (\ref{eqn:tensorstep}) are multiplied
by a damping envelope ${\cal D}(x/x_d)$ which for a Tanh step is \cite{Adshead:2011jq}
\begin{equation}
{\cal D}(y)= \frac{y}{\sinh y}.
\end{equation}

For explicit examples, we first take the sharp step in the potential at
$s_s=\eta_s=3696.9$Mpc with $x_d \gg 1$
 that fits high
frequency oscillations in the Planck temperature power spectrum at high multipole from
Ref.~\cite{Miranda:2013wxa}.  For simplicity we take the limit of a canonical scalar field
$c_s=c_{sa}=1$ and  a tensor-scalar ratio $r=0.2$ after the step as consistent with BICEP2.
Hence
\begin{align}
\epsilon_{Ha} &=\frac{0.2}{16}=0.0125 ,
\end{align}
and for the best fit amplitude
\begin{align}
e_i &=1.254, \nonumber\\
c_b &=c_i = e_b =1 .
\end{align}
Note that for a potential step, $\epsilon_H$ returns to the same value to leading order
well after the step $(e_b=1)$.   

The scalar and tensor power spectrum features
for this model are shown in Fig.~\ref{fig:feature} (left).   Since $c_s=1$, $\eta_s=s_s$ and 
the location of the
features align but the maximum amplitude of the tensor relative to the scalar
features scales as $\epsilon_{Ha}$.  Moreover, the tensor features lack the high $k$ constant
oscillations from $W'$ that make the scalars observable given the decrease in the associated
cosmic variance.   Despite the sharp step in the potential, energy conservation forbids
a sharp step in $H$ and hence unlike the scalars there is no ringing out to high $k$.

In fact,  oscillatory features at high $k$ in tensors are even further suppressed in the observable
CMB B-mode polarization due to the much broader projection of power from $k$ to
multipole $\ell$ associated with B-mode tensor polarization as compared with temperature
or E-mode scalar perturbations \cite{Hu:1997hp}.   Combined these facts imply that
the tensor features associated with this model are too small to be observed.

Next consider a step in the DBI warp that causes a similar step in the quantity $\epsilon_H c_s$ and consequently the scalar power spectrum.    The implied reduction of large scale
scalar power fits the Planck temperature anisotropy data while allowing a large $r=0.2$
tensor contribution to explain the BICEP2 result \cite{Miranda:2014wga}.
For simplicity, we again take $c_{sb}= c_b c_{s a}=1$ and choose parameters
to fit the amplitude and shape of the required reduction
\begin{align}
c_{sa} & =1/c_b= 0.856, \nonumber\\
\epsilon_{Ha} &= \frac{0.2}{16c_{sa}}= 0.0146 ,\nonumber\\
x_d &= 1.43.
\end{align}
Note that this small an $x_d$ does not represent a step that is traversed in much less
than an efold making the sharp step assumption in evaluating GSR integrals only approximate;
however comparisons with the exact scalar calculation show that this approximation suffices
for the description here since even without damping, tensor oscillatory features are suppressed
relative to scalars as we have seen in the previous example.
This model determines the evolution of $c_s$ and $\epsilon_H$ by setting
\begin{align}
c_b &= e_b \approx 1/c_{sa}=1.168 ,\nonumber\\
c_i  &= 0.988, \nonumber\\
e_i &=1.078 .
\end{align}

Fig.~\ref{fig:feature} (right) shows the associated scalar and tensor features.  Again the
overall scale of tensor feature is reduced by $\epsilon_{H}$ compared with the scalar
feature.   In this case, $\epsilon_H$ itself undergoes a step-like change across the feature
and hence the logarithmic rise at small $k$ in the tensor spectrum is due to the change
in tensor tilt in the slow-roll regime before the feature.  However the cosmic variance of
low $k$ modes and the finite size of the current horizon prevents this effect from being
measurable.    In this case the large width of
the step, represented by $x_d$, damps oscillatory features in both scalars and tensors.  
Note that the location of oscillatory features are at slightly smaller $k$ since $\eta_s > s_s$.
Again the combination of these facts imply that tensor features associated with this
model are too small to be observable.      

The common feature of these two examples is that while the ordinary slow-roll approximation
($\epsilon_H \approx $ const.) is strongly violated, slow roll itself is never violated
($\epsilon_H \ll 1$).    Thus they illustrate the fact that large and potentially observable tensor
features requires a fast roll period of inflation.

\section{Discussion} \label{sec:discussion}

In this work, we have developed the generalized slow-roll formalism for tensor power spectrum features from transient  violations of the ordinary slow-roll approximation.   Here the slow-roll parameters are not assumed to be constant and only the background expansion is assumed to
be nearly de Sitter or time translation invariant.   Generally, features
in the scalar power spectrum imply a corresponding set of features in the tensor power
spectrum governed by a generalized consistency relation between their sources.
However, the amplitude of tensor features is suppressed by the slow-roll factor $\epsilon_H$
relative to scalar features and also do not generate as large oscillatory features at 
high wavenumber since the evolution of $H$ is governed by energy conservation and
smoother than its derivatives.

As an illustration of this behavior, we considered inflationary features that are motivated
by anomalies the CMB temperature data, namely high frequency variations at high multipole moment in the Planck data and a step suppression of power at low multipoles.  
Preference for the latter is substantially strengthened by the BICEP2 detection of 
tensor contributions to $B$-mode polarization.  While explanations of either anomaly imply
a matching
set of tensor features, neither require a fast roll period where $\epsilon_H ={\cal O}(1)$ and
in the absence of such a period, tensor features are  greatly suppressed compared with scalar features.

A fast roll period is nonetheless possible if confined to efolds just prior to when the current horizon
exited the horizon during inflation.    For example, these efolds could represent
the end of a prior period of kinetic energy domination \cite{Contaldi:2003zv,Lello:2013awa}.
 Hence the observation of tensor features could provide support
 for such models and distinguish them from slow-roll alternatives that similarly suppress
 large scale temperature power.
 
 More generally, model independent reconstruction of the scalar and tensor source 
 functions $G'_{\curv}$ and $G'_h$ could test the consistency of slow-roll inflation
more generally than the ordinary slow-roll consistency relation \cite{Dvorkin:2011ui}.   Observable deviations from constant $G'_h$
require substantial evolution in the Hubble rate, a violation of time-translation invariance,
 from a relatively fast roll period.   The GSR approach should be useful for calculating 
 the tensor spectrum of such models out to scales approaching the beginning of the
main inflationary period where order unity variations are possible.

 \medskip
 \noindent {\it Acknowledgments}:   WH thanks Peter Adshead for useful conversations and CosKASI where this work was initiated.  This work was supported  by the Kavli Institute for Cosmological Physics at the University of Chicago through grants NSF PHY-1125897 and an endowment from the Kavli Foundation and its founder Fred Kavli and  by U.S.~Dept.\ of Energy contract DE-FG02-13ER41958. 
 
\vfill

\bibliography{Hu14}

\end{document}